\documentclass[preprint,sort&compress, 5p,times]{elsarticle}

\usepackage{lipsum}
\makeatletter
\def\ps@pprintTitle{%
 \let\@oddhead\@empty
 \let\@evenhead\@empty
 \def\@oddfoot{}%
 \let\@evenfoot\@oddfoot}
\makeatother

\usepackage{float,ulem}
\usepackage{graphicx,epsfig,epstopdf,amsmath,amssymb}
\usepackage{xcolor}
\usepackage[colorlinks=true,linkcolor=blue,citecolor=blue,urlcolor=blue]{hyperref}
\usepackage{caption}
\usepackage{subcaption}

\newcommand{\be}{\begin{equation}}
\newcommand{\ee}{\end{equation}}
\newcommand{\ba}{\begin{eqnarray}}
\newcommand{\ea}{\end{eqnarray}}

\def\k{\boldsymbol k}
\begin{document}

\title{Poles from the conserved kinetic equation : The emerging gradient structure and causality riddle of relativistic hydrodynamics}

\author{Sukanya Mitra}
\ead{sukanya.mitra@niser.ac.in}
\address{School of Physical Sciences, National Institute of Science Education and Research, An OCC of Homi Bhabha National Institute, Jatni-752050, India.}

\begin{abstract}
In this work, the poles and the resulting dispersion spectra from the relativistic kinetic equation have been analyzed with the help of a proposed collision kernel that conserves both the energy-momentum tensor and particle current by construction. The dispersion relations, which originally come out in the form of logarithmic divergences, in the long wavelength limit exhibit the systematic gradient structure of the relativistic hydrodynamics. The key result is that, in the derivative expansion series, the spatial gradients appear in perfect unison with the temporal gradients in the non-local relaxation operator like forms. It is then shown that this dispersion structure, including non-local temporal derivatives, is essential for the preservation of causality of the theory truncated at any desired order.
\end{abstract}
\maketitle

\section{Introduction}

In the study of the collective phenomena originating from microscopic dynamics of a system, the kinetic theory \cite{Lifshitz} has proved itself to be quite effective for the underlying coarse-graining method. In order to describe the transport properties of the system, such as momentum and charge diffusion, a description in terms of the weakly coupled quasiparticles and a governing kinetic equation of motion could be applicable at the late times of the emerging macroscopic formalism. The relativistic Boltzmann transport equation \cite{Cercignani,Degroot} has served the purpose quite well \cite{Blaizot:1993zk,Kelly:1994ig,Grozdanov:2016vgg} to extract the characteristic signals of a system and decode its analytic behavior. Since it includes all types of non-analyticity describing both hydrodynamic and non-hydrodynamic sectors, it has found its application both in constructing hydrodynamic formalisms \cite{Romatschke:2009im,Baier:2007ix,Denicol:2010xn} and in the study of
linear response theory through correlation functions \cite{Kovtun:2012rj,Romatschke:2015gic,Hong:2010at,Formanek:2021blc,Bajec:2025dqm,Brants:2024wrx}.

However, the relativistic kinetic equation always had issues treating the highly non-linear microscopic collision term. A suitable approximation comes with a linearized collision kernel, which provides an exact solution in the Fourier space. But the linearization scheme comes with the cost of violating the conservation laws (of the macroscopic field variables) unless some external constraints are imposed in the name of the matching conditions or choice of hydrodynamic frames \cite{Kovtun:2019hdm}. Starting from the Anderson-Witting relaxation time approximation (AWRTA) \cite{AWRTA} to Bhatnagar-Gross-Krook (BGK) method \cite{Bhatnagar:1954zz}, the conservation in terms of both particle current and energy-momentum was never fully realized independent of these constraints. Only recently, \cite{Rocha:2021zcw} presented a fully conserved linearized kernel in terms of the orthogonal  eigenvector basis. In the current work, a collision kernel has been proposed in Eq.\eqref{RTE-2} which is expressed in terms of the out-of-equilibrium field corrections, much in the same line as the BGK kernel, but conserves both particle current and energy-momentum tensor without imposing any hydrodynamic frame choice. It is observed that in order to extract the poles from the conserved kinetic equation and consequently in constructing the dispersion relations, the form of the collision kernel presented here is quite effective.

The dispersion relations are known to characterize the response of the system to perturbations around the equilibrium. In this work, these relations are directly obtained from the solution of the relativistic kinetic equation given in \eqref{RTE-2}, without adopting any particular
hydrodynamic theory. The field corrections used in the collision term to make the conservation possible act as the source term, much in the same line as the external force fields (electromagnetic or gravitational) in the correlator analysis.
Expressing the solution of \eqref{RTE-2} in terms of these fields by taking suitable moments, it is possible to construct a system of simultaneous equations, the determinant of which provides the necessary pole structure that gives rise to the dispersion relations. The pole structure turns out to be in the form of the logarithmic singularity $\Omega_I$, which frequently appears in the expression of the retarded correlators \cite{Romatschke:2015gic,Bajec:2025dqm,Brants:2024wrx} describing the collective behavior of a system.

Next, these dispersion relations are analyzed at the hydrodynamic limit that corresponds to the low energy, long wavelength regime of the spectra. Expansion of the observed pole structures for small wave numbers results in an infinite power series of spatial derivatives resembling the hydrodynamic gradient expansion at near-equilibrium scenario.
The crucial observation is that, for each increasing power of spatial gradients in the series, the temporal derivatives also appear in the denominator, maintaining perfect balance for mode conservation. The condition for mode conservation across a Lorentz boost has been discussed at length in \cite{Hoult:2023clg,Mitra:2024yei,Bhattacharyya:2025hjs}, and it has been argued that the preservation of this condition is imperative for the causality
of the theory. A proper truncation at some desired order of spatial gradient only makes sense if these non-local temporal derivatives are considered appropriately, like raised to a local set of equations. These non-local temporal terms appear in the form of the relaxation operator $(1+i\omega
\tau_R)$ which also sets the radius of convergence of the theory. These operators are actually the signature of the all order derivative corrections of a given causal theory as explained in \cite{Bhattacharyya:2024tfj,Mitra:2023ipl}, and the abrupt truncation of the temporal gradients without treating the whole operator $(1+i\omega
\tau_R)$ directly leads to acausality.

The current analysis depicts how the poles extracted from the relativistic kinetic equation with a fully conserved collision kernel emerge the hydrodynamic gradient expansion at long wave-length limit. Furthermore, the resulting dispersion series includes the in-built mechanism for causality preservation in terms of the non-local temporal operators, indicating the all-order derivative correction for a causal theory. This feature facilitates the claim made in \cite{Mitra:2024yei}, that the underlying microscopic theories are always pathology free, and it is the limitation of the chosen truncation scheme that gives rise to the causality-related issues for a coarse-grained macroscopic theory.

\section{Notations and convention}
Throughout the manuscript, natural unit ($\hbar = c = k_{B} = 1 $) and flat space-time with mostly negative metric signature $g^{\mu\nu} = \text{diag}\left(1,-1,-1,-1\right)$ have been used. $n, \varepsilon, P, T$ and $\mu$ respectively denote the particle number density, energy density, pressure, temperature, and chemical potential. The particle momenta is denoted by $p^{\mu}$ with single particle energy $E_p=p^0=p^{\mu}u_{\mu}$, where the fluid velocity $u^{\mu}$ in the global rest frame is given by $(1,0,0,0)$. $x^{\mu}$ is the space-time coordinate and $\partial_{\mu}$ defines the covariant derivative. $d\Gamma_p=\frac{d^3p}{(2\pi)^3 p^0}$ is the phase-space factor for moment integral. The Fourier transformation considered here is defined as $\psi(x) \sim
\int d^4 k ~ e^{ik_{\mu}x^{\mu}} \tilde{\psi}(k)$, with the wave 4-vector $k^{\mu}=(\omega,0,0,k)$, following $\partial_{\mu}=ik_{\mu}$.
$\eta$ and $D_F$ are the coefficients of shear viscosity and charge diffusion respectively.
$\tau_R$ is the relaxation time for the single particle distribution function.

\section{Moment definitions}
The fundamental fluid fields, namely the particle 4-current ($J^{\mu}$) and the energy-momentum tensor ($T^{\mu\nu}$) can be expressed as the moments folded over the single particle distribution function $f(x,p)$ respectively as follows,
\begin{align}
 J^{\mu}(x)=\int d\Gamma_p p^{\mu} f(x,p)~,~~~
 T^{\mu\nu}(x)=\int d\Gamma_p p^{\mu}p^{\nu} f(x,p)~.
 \label{fields}
\end{align}
Apart from that, two additional moments are defined here over the local equilibrium distribution function $f_{eq} (x,p)$ as,
\begin{align}
 a_n=&\frac{1}{T^n} \int d\Gamma_p f_{eq}(x,p)E_p^n~,
 \label{moma}\\
 \Delta^{\mu\nu}b_n=&\frac{1}{T^{n+2}} \int d\Gamma_p f_{eq}(x,p)p^{\langle\mu\rangle}p^{\langle\nu\rangle}E_p^n~,
 \label{momb}
\end{align}
with $p^{\langle\mu\rangle}=\Delta^{\mu\nu}p_{\nu}$ and $\Delta^{\mu\nu}=g^{\mu\nu}-u^{\mu}u^{\nu}$.
Further, for the analysis of poles from the kinetic equation, the following three moments are defined in the $k^{\mu}$ space over the global equilibrium $f_0(p)$,
\begin{align}
 &I_m(\omega,k)=\frac{i}{\tau_R}\int d\Gamma_p~f_0(p)~\frac{\left(p^{\mu}u_{\mu}\right)^m}{\left[-\left(p^{\mu}k_{\mu}\right)+\frac{i}{\tau_R}\left(p^{\mu}u_{\mu}\right)\right]}~,\label{momI}\\
 &J_m^{\mu}(\omega,k)=\frac{i}{\tau_R}\int d\Gamma_p~f_0(p)~\frac{\left(p^{\mu}u_{\mu}\right)^m p^{\langle\mu\rangle}}{\left[-\left(p^{\mu}k_{\mu}\right)+\frac{i}{\tau_R}\left(p^{\mu}u_{\mu}\right)\right]}~,\label{momJ}
 \\
 &Q_m^{\mu\nu}(\omega,k)=\frac{i}{\tau_R}\int d\Gamma_p~f_0(p)~\frac{\left(p^{\mu}u_{\mu}\right)^m p^{\langle\mu\rangle} p^{\langle\nu\rangle}}{\left[-\left(p^{\mu}k_{\mu}\right)+\frac{i}{\tau_R}\left(p^{\mu}u_{\mu}\right)\right]}~.
 \label{momQ}
\end{align}
It can be shown that with the chosen direction of the wave 4-vector $k^{\mu}=(\omega,0,0,k)$, the only surviving components of the above vector and tensor integrals are $J_m^z, Q_m^{zz}, Q_m^{xx}, Q_m^{yy}$.

\section{The proposed collision kernel and the conservation laws}
The relativistic transport equation (which actually is the well known covariant Boltzmann equation), in absence of any force or mean-field interaction term is given as the following \cite{Cercignani,Degroot},
\begin{align}
p^{\mu}\partial_{\mu}f(x,p)=C[f,f]~.
\label{RTE-1}
\end{align}
Eq.\eqref{RTE-1} describes the evolution of the single particle distribution function $f$ in terms of the collision term $C$ that encodes the system interaction.
$f(x,p)$ is the out-of-equilibrium distribution function that deviates from the local equilibrium distribution $f_{eq}(x,p)$ in the following manner,
\begin{align}
 f(x,p)=f_{eq}(x,p)+\Delta f(x,p)~,~~\Delta f=f_{eq}(x,p)~\phi(x,p)~,
 \label{dev1}
\end{align}
with $\Delta f(x,p)$ (as well as $\phi(x,p)$) being the out-of-equilibrium distribution deviation from the local equilibrium.
$C[f,f]$ in general is a 9-dimensional momentum integral folded over the scattering cross-section of particle interaction (the typical detailed form can be found in \cite{Degroot}).

For the practical purpose of analytical treatment of the integro-differential Eq.\eqref{RTE-1}, the collision kernel $C[f,f]$ is required to be linearized over the deviation function $\phi$. One of the attempts was the AWRTA model that describes the transport equation with the simplest possible collision kernel as $p^{\mu}\partial_{\mu}f=-\frac{p^{\mu}u_{\mu}}{\tau_R}f_{eq}\phi$ \cite{AWRTA}. $\tau_R$ is called the relaxation time - the timescale within which the system restores the equilibrium. Although widely used in the applications of kinetic equation and qualitatively effective in describing transport properties \cite{Denicol:vaa}, this approximation of the collision operator is blighted by the fundamental conservation issues unless some macroscopic constraints (we call them matching condition or the choice of hydrodynamic frames such as Landau frame in this case) are externally imposed. An advancement can be made by using the BGK collision kernel \cite{Bhatnagar:1954zz} which conserves the particle four-current by construction from Eq.\eqref{RTE-1}. However, the BGK collision kernel still lacks the energy-momentum conservation. There are recent attempts to incorporate the energy-momentum conservation in some modified BGK formalism \cite{Singha:2023eia}, but in order to realize both current and momentum conservation, at least some frame constraint is necessary. These frame choices are basically the out- of-equilibrium definition of the dynamical variables (such as $T,\mu,u^{\mu}$). As very nicely put in \cite{Kovtun:2019hdm}, the number of such choices could be infinity (unless the physical criteria for the underlying theory need to eliminate some of them) and limiting to any specific one duly cost the generality. In such a scenario, a BGK like collision kernel that respects both current and momentum conservation by construction without requiring any such external imposition is essential for the theoretical analysis of the kinetic equation \eqref{RTE-1}. In light of the above discussion, I am proposing here the following collision kernel that conserves both the particle 4-current $J^{\mu}(x)$ and energy-momentum tensor $T^{\mu\nu}(x)$ by construction as,
\begin{align}
 p^{\mu}\partial_{\mu}f(x,p)=&-\frac{p^{\mu}u_{\mu}}{\tau_R}f_{eq}(x,p)~\Bigg[\phi-
 \frac{1}{\Delta_a}\left\{\frac{a_3}{T}\Delta n-\frac{a_2}{T^2}\Delta\varepsilon\right\}\nonumber\\
 &~~+E_p\frac{1}{\Delta_a}\left\{\frac{a_2}{T^2}\Delta n-\frac{a_1}{T^3}\Delta\varepsilon\right\}-\frac{p^{\langle\mu\rangle}\Delta W_{\mu}}{b_1T^3}\Bigg]~,
 \label{RTE-2}
\end{align}
with $\Delta_a=\left(a_1a_3-a_2^2\right)$. In the linearized collision term, the used dissipative field corrections for particle number density, energy density and vector heat current are respectively given by,
\begin{align}\label{diss}
&\Delta n(x)=\int d\Gamma_p \left(p^{\mu}u_{\mu}\right)\Delta f~,~~~~\Delta \varepsilon(x)=\int d\Gamma_p \left(p^{\mu}u_{\mu}\right)^2\Delta f~,\nonumber\\
&~~~~~~~~~~~~~~~\Delta W^{\alpha}(x)=\int d\Gamma_p \left(p^{\mu}u_{\mu}\right)p^{\langle\alpha\rangle}\Delta f~.
\end{align}
Taking the zeroth and first moment of Eq.\eqref{RTE-2}, one can readily see that the right hand sides of both the moment equations identically vanish. Following the field definitions given in Eq.\eqref{fields}, this immediately leads to the following conservation laws,
\begin{align}
 \partial_{\mu}J^{\mu}=0~,~~~~~~~~\partial_{\mu}T^{\mu\nu}=0~.
\end{align}
One thing needs to be mentioned here. Relaxation time approximation in relativistic Boltzmann
equation which is compatible with both micro and macroscopic conservation laws, has been extensively studied in \cite{Rocha:2021zcw}. A slight different version of the same (with a different momentum basis) has been presented in \cite{Biswas:2022cla}. It is not claimed here that the collision term given in Eq.\eqref{RTE-2} is entirely novel in that sense.
The importance of the form given in \eqref{RTE-2} is that the collision kernel introduced here is duly expressed in terms of the macroscopic field variables much in the same line as the BGK. The last three terms of the collision kernel are essentially the zero modes (coefficients of $1, E_p$ and $p^{\langle\mu
\rangle}$, more details can be found in \cite{Biswas:2022cla}) that make the conservation possible. Expressing them entirely as the functions of dissipative field corrections gives us the liberty to treat the kinetic equation in terms of the macroscopic variables. We will see in the coming sections that this feature is required for the further treatment of the conserved transport equation.

\section{Poles from the relativistic kinetic equation}
To proceed further, let us first define the field perturbations around the global equilibrium. Following the treatment opted in \cite{Bajec:2025dqm}, the shifts respectively in the out-of-equilibrium distribution $f(x,p)$ and the local equilibrium distribution $f_{eq}(x,p)$, from the global equilibrium
function $f_0(p)$ are defined as,
\begin{equation}
 f(x,p)=f_0(p)+\delta f(x,p)~,~~~~
 f_{eq}(x,p)=f_0(p)+\delta f_{eq}(x,p).
 \label{dev2}
\end{equation}
Recalling Eq.\eqref{dev1} and comparing with \eqref{dev2} one can trivially see,
\begin{equation}
\Delta f(x,p)=\delta f(x,p)-\delta f_{eq}(x,p)~,
\label{pert1}
\end{equation}
as well as from Eq.\eqref{diss},
\begin{align}
\Delta n=&\delta  n -\delta n_{eq}~,~~~
\Delta \varepsilon=\delta \varepsilon-\delta \varepsilon_{eq}~,~~~\Delta W^{\alpha}=W^{\alpha}-W_{eq}^{\alpha},
\label{pert2}\\
 \text{with,}~~&\left\{\delta n,~\delta n_{eq}\right\}=\int d\Gamma_p \left(p^{\mu}u_{\mu}\right)~\left\{\delta f,~\delta f_{eq}\right\}~,\nonumber\\
 &\left\{\delta\varepsilon,~\delta\varepsilon_{eq}\right\}=\int d\Gamma_p \left(p^{\mu}u_{\mu}\right)^2~\left\{\delta f,~\delta f_{eq}\right\}~,\nonumber\\
 &\left\{W^{\alpha},~W^{\alpha}_{eq}\right\}=\int d\Gamma_p \left(p^{\mu}u_{\mu}\right)p^{\langle\alpha\rangle}~\left\{\delta f,~\delta f_{eq}\right\}~.
 \label{diss2}
\end{align}
In analyses like \cite{Bajec:2025dqm,Romatschke:2015gic}, the quantities $\delta\psi$ and $\delta\psi_{eq}$ have been treated on the same footing, which is absolutely perfect for an AWRTA collision kernel and the declared Landau matching condition (Eq.(13) of \cite{Bajec:2025dqm}). However, in a general hydrodynamic frame (where the dissipative corrections $\Delta\psi~(=\psi(x)-\psi_{eq}(x))$ are non-zero) with the collision term \eqref{RTE-2} this liberty is not available anymore. First, we consider the local equilibrium distribution function as, $f_{eq}(x,p)=\text{exp}\left[-\frac{p_{\mu}u_{eq}^{\mu}(x)}{T_{eq}(x)}+\frac{\mu_{eq}(x)}{T_{eq}(x)}\right]$, with the corresponding local fields around global equilibrium (denoted by subscript zero) as $\psi_{eq}(x)=\psi_0+\delta\psi_{eq}(x)$, where $\delta \psi_{eq}$ is the global to local equilibrium field correction. Then keeping up to linear terms in perturbations we find that,
\begin{equation}
\label{pert3}
 \delta f_{eq}(x,p)=f_0\left[\frac{\delta\mu_{eq}}{T_0}-\frac{p_{\mu}\delta u_{eq}^{\mu}}{T_0}+\left\{\frac{p^0}{T_0}-\frac{\mu_0}{T_0}\right\}
 \frac{\delta T_{eq}}{T_0}\right]~,
\end{equation}
with $f_0(p)=\text{exp}\left[-\frac{p^0}{T_0}+\frac{\mu_0}{T_0}\right]$. Consequently, we have,
\begin{align}\label{pert4}
&\delta n_{eq}=a_1~\delta\mu_{eq}+a_2~\delta T_{eq}-a_1~\mu_0~\delta T_{eq}/T_0~,\nonumber\\
&\delta\varepsilon_{eq}=a_2~T_0~\delta\mu_{eq}+a_3~T_0~\delta T_{eq}-a_2~\mu_0~\delta T_{eq}~,\nonumber\\
&W^{\alpha}_{eq}=-b_1~(T_0)^2~\delta u_{eq}^{\alpha}~.
\end{align}
Here the property that $\delta u^{\mu}_{eq}$ is orthogonal to $u^{\mu}_0$ (since velocity normalization requires $u_{0\mu}\delta u^{\mu}_{eq}={\cal{O}}(\delta^2)$) has been used.
Putting \eqref{pert3} and \eqref{pert4} in Eq.\eqref{RTE-2} and doing the required algebra, we see that the dissipative field corrections $\Delta\psi$ (local to out-of-equilibrium) in the collision kernel is replaced by the field correction $\delta\psi(x)=\psi(x)-\psi_0$ (global to out-of-equilibrium) in the following manner,
\begin{align}
 &p^{\mu}\partial_{\mu}\Big\{\delta f(x,p)\Big\}=-\frac{E_p}{\tau_R}\Bigg[\delta f(x,p)-f_0~
 \frac{1}{\Delta_a}\left\{\frac{a_3}{T}\delta n(x)-\frac{a_2}{T^2}\delta\varepsilon(x)\right\}\nonumber\\
 &~~~~~~+f_0~E_p\frac{1}{\Delta_a}\left\{\frac{a_2}{T^2}\delta n(x)-\frac{a_1}{T^3}\delta\varepsilon(x)\right\}-f_0~\frac{p^{\langle\mu\rangle}W_{\mu}(x)}{b_1T^3}\Bigg]~.
 \label{RTE-3}
\end{align}
Next, we solve Eq.\eqref{RTE-3} into the Fourier space (convention mentioned earlier) and obtain the following solution for $\tilde{\delta f}(k,p)$ as,
\begin{align}\label{deltaf}
 &\tilde{\delta f}(k,p)=\frac{i}{\tau_R}~f_0(p)~\times\nonumber\\
 &\frac{\Bigg[\frac{E_p}{\Delta_a}\Big\{\frac{a_3}{T}\tilde{\delta n}(k)-
 \frac{a_2}{T^2}\tilde{\delta\varepsilon}(k)\Big\}-\frac{E_p^2}{\Delta_a}\Big\{\frac{a_2}{T^2}\tilde{\delta n}(k)-\frac{a_1}{T^3}\tilde{\delta\varepsilon}(k)\Big\}+\frac{E_pp^{\langle\mu\rangle}}{b_1T^3}\tilde{W}_{\mu}(k)\Bigg]}{\left\{-p_{\mu}k^{\mu}+i\frac{E_p}{\tau_R}\right\}}.
\end{align}
To avoid clutter the scaling temperature notation is still kept $T$ (instead of $T_0$) since for linear perturbations that anyway does not matter.

The distribution function (actually it's correction) in Eq.\eqref{deltaf} is now ready for further treatment. Following Eq.\eqref{diss2}, the field corrections $\tilde{\delta n}, \tilde{\delta\varepsilon}$ and $\tilde{W}^{\alpha}$
can be obtained from \eqref{deltaf} which leads to a system of simultaneous equations given below,
\begin{equation}
\label{matrix}
 AX=0~, ~~~~~~X=\left[\tilde{\delta n},\tilde{\delta\varepsilon},\tilde{W}^z,\tilde{W}^x,\tilde{W}^y\right]^T~,
\end{equation}
with,
$$A=
\begin{bmatrix}
B_{(3\times 3)} & 0 & 0\\
0 & \left\{1+\frac{Q_2^{xx}}{b_1T^3}\right\} & 0\\
0 & 0 & \left\{1+\frac{Q_2^{yy}}{b_1T^3}\right\}
\end{bmatrix}~,
$$
which includes,
$$B=
\begin{bmatrix}
\left\{1-\frac{a_3I_2}{T\Delta_a}+\frac{a_2I_3}{T^2\Delta_a}\right\} & \left\{\frac{a_2I_2}{T^2\Delta_a}-\frac{a_1 I_3}{T^3\Delta_a}\right\} & \left\{\frac{J_2^z}{b_1T^3}\right\} \\
\left\{\frac{a_2I_4}{T^2\Delta_a}-\frac{a_3I_3}{T\Delta_a}\right\} & \left\{1+\frac{a_2I_3}{T^2\Delta_a}-\frac{a_1I_4}{T^3\Delta_a}\right\} & \left\{\frac{J_3^z}{b_1T^3}\right\} \\
\left\{\frac{a_2J_3^z}{T^2\Delta_a}-\frac{a_3J_2^z}{T\Delta_a}\right\} & \left\{\frac{a_2J_2^z}{T^2\Delta_a}-\frac{a_1J_3^z}{T^3\Delta_a}\right\} & \left\{1+\frac{Q_2^{zz}}{b_1T^3}\right\}
\end{bmatrix}~.
$$
For nontrivial solutions of Eq.\eqref{matrix}, the determinant of $A$ must vanish which leads to the poles that finally give rise to the dispersion structure. The three poles thus obtained from Eq.\eqref{matrix} are listed below as follows,
\begin{align}
 \Delta^{x}=&1+\frac{Q_2^{xx}}{b_1T^3}=0~,~~~~~~
  \Delta^{y}=1+\frac{Q_2^{yy}}{b_1T^3}=0~,
  \label{pole1}
  \\
  \Delta^z=&\left(1+\frac{Q_2^{zz}}{b_1T^3}\right)\left\{1+\frac{1}{\Delta_a}\left(\frac{2a_2}{T^2}I_3-\frac{a_1}{T^3}I_4-\frac{a_3}{T}I_2+\frac{I_2I_4}{T^4}-\frac{I_3^2}{T^4}\right)\right\}\nonumber\\
  &+\frac{\left(J_2^z\right)^2}{b_1T^3}\frac{1}{\Delta_a}\left\{\frac{a_3}{T}-\frac{I_4}{T^4}\right\}+\frac{\left(J_3^z\right)^2}{b_1T^3}\frac{1}{\Delta_a}\left\{\frac{a_1}{T^3}-\frac{I_2}{T^4}\right\}\nonumber\\
  &-2\frac{J_2^zJ_3^z}{b_1T^3}\frac{1}{\Delta_a}\left\{\frac{a_2}{T^2}-\frac{I_3}{T^4}\right\}~=~0~.
  \label{pole2}
\end{align}

\section{Analysis of poles : The logarithmic structure}
The remaining task is to simplify the integrals $I_m, J_m^{\mu}$ and $Q_m^{\mu\nu}$ in order to express $\Delta^{x,y,z}$ from Eq.\eqref{pole1} and \eqref{pole2} in terms of the frequency $\omega$ and wave-vector $k$. From Eq.\eqref{momI}, \eqref{momJ} and \eqref{momQ}, it is possible to separate the angular integrals (that exclusively include the $\omega$ and $k$ dependence) in the following way,
\begin{align}
\label{momfin}
 I_m&=\frac{i}{\tau_R}\Omega_I\int\frac{d|\vec{p}||\vec{p}|^2}{2\pi^2}\left(E_p\right)^{m-2}f_0~,\nonumber\\
 J_m^z&=\frac{i}{\tau_R}\Omega_J\int\frac{d|\vec{p}||\vec{p}|^3}{2\pi^2}\left(E_p\right)^{m-2}f_0~,~~~J_m^x=J_m^y=0~,\nonumber\\
 Q_m^{zz}&=\frac{i}{\tau_R}\Omega_Q^z\int\frac{d|\vec{p}||\vec{p}|^4}{2\pi^2}\left(E_p\right)^{m-2}f_0~,~~~Q_m^{xz}=Q_m^{yz}=0~,\nonumber\\
 Q_m^{xx}&=Q_m^{yy}=\frac{i}{\tau_R}\Omega_Q^{x,y}\int\frac{d|\vec{p}||\vec{p}|^4}{2\pi^2}\left(E_p\right)^{m-2}f_0~,~~~Q_m^{xy}=0~,
\end{align}
where,
\begin{align}
\label{ang}
 &\Omega_I=\int\frac{d\Omega}{4\pi}\frac{1}{\left[\frac{i}{\tau_R}-\omega+\varv_p k ~\text{cos}(\theta)\right]}=\frac{1}{2\varv_p k}\text{log}\left\{\frac{a-1}{a+1}\right\},\nonumber\\
 &\Omega_J=\frac{1}{\varv_pk}+a\Omega_I~,~~~~\Omega_Q^z=a\Omega_J~,~~~~\Omega_Q^{x,y}=\frac{1}{2}\Omega_I-\frac{1}{2}\Omega_Q^z~,\nonumber\\
 &\text{with,}~~~ a=\frac{1}{\varv_p}\frac{(1+i\omega\tau_R)}{(ik\tau_R)}, ~~~~~\text{and}~~~\varv_p=\frac{|\vec{p}|}{p^0}~.
\end{align}
To further simplify the results, the ultra-relativistic scenario is considered here, for which one has $|\vec{p}|=p^0,~\varv_p=1$ and $b_n=-\frac{1}{3}a_{n+2}$. One can proceed with the massive case where the momentum integrals in \eqref{momfin} will be expressed in terms of the modified Bessel function of second kind instead of Gamma functions but the basic functional form of the angular integrals in \eqref{ang} will not alter.

The analysis leads us to three separate pole structures. The first two poles are contributed solely from the components of $\tilde{W}^{\mu}$  that are perpendicular to the wave-vector direction (i.e., $\tilde{W}^x$ and $\tilde{W}^y$) and are given by,
\begin{equation}
 \Delta^x=\Delta^y=1-3\frac{i}{\tau_R}\Omega_Q^{x,y}=0~.
 \label{poleshear}
\end{equation}
These are clearly the two shear or transverse modes since $W^{\mu}$ by construction is perpendicular to $u^{\mu}$ and hence in the same direction as the velocity fluctuation $\delta u^{\mu}$ ($\delta u^{\mu}$ must be normal to $u^{\mu}$ to maintain velocity normalization). The perturbation corresponding to the fluid flow perpendicular to the gradient direction (direction of $\vec{k}$) is the well known shear mode of the fluid. The third pole results combinedly from the perturbations $\tilde{\delta n},~ \tilde{\delta\varepsilon}$ and $\tilde{W}^z$ and is given by,
\begin{equation}
 \Delta^z=\left(1-\frac{i}{\tau_R}\Omega_I\right)\left\{\left(1-\frac{i}{\tau_R}\Omega_I\right)\left(1-3\frac{i}{\tau_R}\Omega_Q^z\right)-3\left(\frac{i}{\tau_R}\Omega_J\right)^2\right\}=0~.
 \label{polediffsound}
\end{equation}
We can observe that Eq.\eqref{polediffsound} includes two individual poles. The first one,
\begin{equation}
 1-\frac{i}{\tau_R}\Omega_I=0~,
 \label{polediff}
\end{equation}
corresponds to the diffusion channel. The second one,
\begin{equation}
 \left(1-\frac{i}{\tau_R}\Omega_I\right)\left(1-3\frac{i}{\tau_R}\Omega_Q^z\right)-3\left(\frac{i}{\tau_R}\Omega_J\right)^2=0~,
 \label{polesound}
\end{equation}
is the longitudinal or the sound channel.
Eq.\eqref{poleshear}, \eqref{polediff} and \eqref{polesound} are the four poles or the dispersion relations obtained from the conserved relativistic transport equation and are the first set of main results of the current analysis.

It can be observed from Eq.\eqref{ang} that the spectral functions $F(\omega,k)=0$ given in \eqref{poleshear}, \eqref{polediff} and \eqref{polesound}, have logarithmic dependence via the angular integrals $\Omega_{I}, \Omega_{J}$ and $\Omega_{Q}$. These types of logarithmic divergences are found to be quite common in the collective mode analysis of the effective kinetic theory description based on the Boltzmann equation \cite{Kovtun:2012rj,Brants:2024wrx}. In fact, the quantity $\Omega_I$ frequently appears in the studies of the retarded correlation functions \cite{Romatschke:2015gic,Bajec:2025dqm}. The analytic structure of this logarithmic function and its connection to causality have been discussed in details in section \ref{caus}.

\section{Hydrodynamic limit : The emerging gradient structure}
Next, we want to analyze the results obtained in the last section in the hydrodynamic limit. The hydrodynamic limit of a given theory corresponds to the small values of frequencies and wave numbers. Expanding the poles $\Delta^{x,y,z}=0$ in the powers of $\omega$ and $k$ at that limit, one can extract the corresponding dispersion relations. The expansion formula used for the complex logarithmic function is given below,
\begin{equation}
\label{logex1}
 \text{log}(z)=2\left[\left(\frac{z-1}{z+1}\right)+\frac{1}{3}\left(\frac{z-1}{z+1}\right)^3+\frac{1}{5}\left(\frac{z-1}{z+1}\right)^5+\cdots\right]~,~{\cal{R}}(z)>0.
\end{equation}
Eq.\eqref{logex1} perfectly holds for small $k$ limit and leads to the following expansion of the logarithmic function obtained in \eqref{ang},
\begin{equation}
\label{logex2}
 \text{log}\left(\frac{a-1}{a+1}\right)=-2\left\{\frac{1}{a}+\frac{1}{3}\left(\frac{1}{a}\right)^3+\frac{1}{5}\left(\frac{1}{a}\right)^5+\cdots\right\}~.
\end{equation}
Here we note that, since $\frac{1}{a}=(ik\tau_R)/(1+i\omega\tau_R)$, the expansion given in \eqref{logex2} can be treated as a perturbative series in the small $k$ limit.

Following the above prescription, we first analyze the two shear poles given in \eqref{poleshear}. Using expansion \eqref{logex2}, Eq.\eqref{poleshear} reduces to the dispersion relation given below,
\begin{equation}\label{dispshear}
 \left(i\omega\tau_R\right)=\frac{1}{5}~\frac{\left(ik\tau_R\right)^2} {\left(1+i\omega\tau_R\right)^2}+\frac{3}{35}~\frac{\left(ik\tau_R\right)^4}{ \left(1+i\omega\tau_R\right)^4}+\cdots ~.
\end{equation}
By similar treatment, the diffusion channel pole of $\Delta^z=0$ from Eq.\eqref{polediff} gives the following dispersion relation,
\begin{equation}\label{dispdiff}
 \left(i\omega\tau_R\right)=\frac{1}{3}~\frac{\left(ik\tau_R\right)^2} {\left(1+i\omega\tau_R\right)^2}+\frac{1}{5}~\frac{\left(ik\tau_R\right)^4}{ \left(1+i\omega\tau_R\right)^4}+\cdots ~.
\end{equation}
Finally, the expansion of the sound pole from Eq.\eqref{polesound} in the small wave-number limit derives a bit involved expression for the dispersion relation as the following,
\begin{align}\label{dispsound}
 \frac{1}{3}~\left(i\omega\tau_R\right)^2=\frac{1}{5}~\left(ik\tau_R\right)^2\left\{\frac{14}{9}~\frac{1}{\left(1+i\omega\tau_R\right)}-\frac{1}{\left(1+i\omega\tau_R\right)^2}\right\}\nonumber\\+\frac{1}{7}~\left(ik\tau_R\right)^4\left\{\frac{22}{15}~\frac{1}{\left(1+i\omega\tau_R\right)^3}-\frac{1}{\left(1+i\omega\tau_R\right)^4}\right\}+\cdots ~.
\end{align}
The dispersion relations given in \eqref{dispshear}, \eqref{dispdiff}, and \eqref{dispsound} are the second set of results of the current analysis, where we see that the right hand sides of the spectral relation $\omega(k)$ are the infinite power series over the small wave number $k$. Remembering that the $k$ powers are generated in the Fourier analysis from the spatial derivatives, these dispersion series are equivalent to the gradient expansion of relativistic hydrodynamics in near equilibrium scenario. Akin to the hydrodynamic theory, where order truncation is regularly done in terms of the spatial gradients, these dispersion series can also be truncated at any desirable order in power $k$.

Here comes a significant observation of the current analysis. In Eq.\eqref{dispshear}, \eqref{dispdiff} and \eqref{dispsound}, the infinite sum not only involves the spatial derivative $k$, but also the temporal derivative $\omega$ in a relaxation-operator like form $(1+i\omega\tau_R)$ in the denominator of each $k$ powered term of the series. From here on, such terms that include powers of $\frac{1}{\left(1+i\omega\tau_R\right)}$ (where the time derivative is appearing in the denominator) will be called as the ``non-local'' terms or ``non-local'' operators indicating
a nonlocality in time (or integration over time). The power over these non-local operators
also systematically increases for each higher $k$ powered term in the series. This structure of competing spatial and temporal derivatives is also observed in \cite{Mitra:2024yei} and has profound effects on the causality of the theory. The dispersion series \eqref{dispshear}, \eqref{dispdiff} and \eqref{dispsound} can be truncated at any $k$ powered term, as long as the factor $(1+i\omega\tau_R)$ is appearing in the denominator with appropriate power, such that the number of modes are conserved across an arbitrary Lorentz boost. The problem of retaining such non-local terms in any gradient expansion is that such series always bears the risk of divergence, especially near the radius of convergence of the theory ($\omega\approx \frac{i}{\tau_R}$). As a remedy, these non-local factor $(1+i\omega\tau_R)$ can be lifted in the left hand side to recast the dispersion polynomial into a local set of equations. The crucial point is that, without this factor the causality of the theory is severely impeded. The reason is that, without the factor $(1+i\omega\tau_R)$ in the denominator, the power of $k$ always exceeds the power of $\omega$ for any given truncation in order $k$. This actually violates the mode conservation condition and directly leads to acausality of the system of equations \cite{Hoult:2023clg,Bhattacharyya:2025hjs}. This factor in the denominator actually indicates the all-order resummation of temporal derivations in a causal theory \cite{Bhattacharyya:2024tfj,Mitra:2023ipl} and for the same reason the abrupt truncation in the temporal derivations in form $(1+i\omega\tau_R)^{-1}=1+(i\omega\tau_R)+(i\omega\tau_R)^2+\cdots$ directly leads to pathology (more discussion can be found in \cite{Mitra:2024yei}). The only exception is observed to be the sound channel leading term in \eqref{dispsound} and will be explained later why.

Next, it will be shown how the dispersion polynomials \eqref{dispshear}, \eqref{dispdiff} and \eqref{dispsound} behave after a certain truncation in $k$ power. Keeping terms up to ${\cal{O}}(k^2)$,
Eq.\eqref{dispshear} becomes,
\begin{equation}\label{polyshear}
 \left(i\omega\tau_R\right)\left(1+i\omega\tau_R\right)^2-\frac{1}{5}\left(ik\tau_R\right)^2=0~.
\end{equation}
At small $k$ limit, one way to decompose \eqref{polyshear} is as follows,
\begin{align}\label{modeshear}
 \omega=\frac{i}{\tau_R}~,~~~~~~\tau_R\omega^2-i\omega-\frac{\eta}{\varepsilon_0+P_0} k^2=0~,
\end{align}
with $\frac{\eta}{\varepsilon_0+P_0}=\frac{\tau_R}{5}$ \cite{Mitra:2020gdk}.
The first solution of \eqref{modeshear} in a pure non-hydro, non-propagating mode. The second solution can be recognized as the well known M{\"u}ller-Israel-Stewart (MIS) shear modes of relativistic hydro \cite{Pu:2009fj}. In a very similar treatment, the truncation of \eqref{dispdiff} up to ${\cal{O}}(k^2)$ can also be decomposed at small $k$ limit as,
\begin{align}\label{modediff}
 \omega=\frac{i}{\tau_R}~,~~~~~~\tau_R\omega^2-i\omega-D_F~ k^2=0~,
\end{align}
with $D_F=\frac{\tau_R}{3}$ \cite{Romatschke:2015gic}. \eqref{modediff} can also be identified as a pure non-hydro, non-propagating mode and the MIS diffusion modes \cite{Brito:2020nou} respectively. Finally, truncating Eq.\eqref{dispsound} up to order ${\cal{O}}(k^2)$, we have the following dispersion polynomial,
\begin{equation}\label{polysound}
 \left(i\omega\tau_R\right)^2 \left(1+i\omega\tau_R\right)^2-\frac{14}{15}\left(i\omega\tau_R\right)\left(ik\tau_R\right)^2-\frac{1}{3}\left(ik\tau_R\right)^2=0~.
\end{equation}
A similar decomposition of \eqref{polysound} at small $k$ limit renders the following solutions,
\begin{align}
\label{modesound}
 &\omega=\frac{i}{\tau_R}~,\nonumber\\
 &\tau_R\omega^3-i\omega^2-\left\{\frac{\tau_R}{3}+\frac{4}{3}\frac{\eta}{(\varepsilon_0+P_0)}\right\}\omega k^2+\frac{i}{3}k^2=0~.
\end{align}
The first solution of \eqref{modesound} is the same non-hydro, non-propagating mode like the previous cases. The second solution, however, can be identified as the known sound mode of the MIS theory \cite{Denicol:2008ha} with $\frac{\eta}{\varepsilon_0+P_0}=\frac{\tau_R}{5}$. The individual mode solutions from each of \eqref{modeshear}, \eqref{modediff} and \eqref{modesound} (both hydro and non-hydro) include the transport parameters like $\tau_R, \eta, D_F$ apart from the leading term of the hydrodynamic mode of \eqref{modesound} which is $\omega=\pm\frac{k}{\sqrt{3}}+{\cal{O}}(k^2)$. This leading term belongs to the ideal fluid. One can observe that, if the $1/(1+i\omega\tau_R)$ factors on the leading $k$ expansion term (${\cal{O}}(k^2)$ ) in \eqref{dispsound} is replaced by $1$ (basically the leading term in small $\omega$ expansion), we obtain the same ideal mode $\omega=\pm\frac{k}{\sqrt{3}}$. This means, whenever relativistic theories include dissipation the infinitely resummed non-local factor $(1+i\omega\tau_R)$ is bound to appear in the denominator to maintain causality. This non-local set of equations can then be recast into a local set of equations
by `integrating in' new `non­-fluid' degrees
of freedom apart from the fundamental fluid fields like temperature and velocity \cite{Mitra:2024yei,Bhattacharyya:2024tfj}.
In a way, that is how the underlying microscopic theory keeps the non-hydrodynamic signature alive in the resulting hydrodynamic description. It is discussed in further details in section \ref{caus}.

\section{Preservation of causality}
\label{caus}
We observe that the key ingredient of the pole structure of relativistic kinetic equation is the following logarithmic function,
\be\label{logsing}
\text{log}\left\{\frac{a-1}{a+1}\right\}=\text{log}\frac{\left[\omega-k-\frac{i}{\tau_R}\right]}{\left[\omega+k-\frac{i}{\tau_R}\right]}~,
\ee
resulting from the angular integrals given in \eqref{ang}. Eq.\eqref{logsing} includes logarithmic branch cuts origination from the singularities at $\omega=\frac{i}{\tau_R}\pm k$.
These cuts are the typical non-hydrodynamic signature of the kinetic theory. In fact this
analytic structure is a direct consequence of the RTA Boltzmann equation cut, starting at $\omega=\frac{i}{\tau_R}\pm k$. These non-hydrodynamic properties are the indicators of the causality in any given theory. As very nicely remarked in \cite{Bemfica:2017wps}, the underlying microscopic descriptions of a hydro theory such as kinetic theory are always pathology free. So, the preservation of causality of a hydrodynamic formulation derived from such kinetic theory requires the inclusion of these non-hydro informations into the resulting coarse-grained theory \cite{Hoult:2023clg,Heller:2022ejw}. The treatment of the kinetic poles at hydrodynamic limit renders the logarithmic divergences to simple pole structure, albeit the price is paid by an all order gradient correction. The resulting gradient series in Eq.\eqref{dispshear}, \eqref{dispdiff} and \eqref{dispsound} bear the non-hydro signature in the form of the non-local operators. A look at this operators will reveal that they are infinite sum over the temporal derivatives in the form $\sum_{n=0}^{\infty}(-i\omega\tau_R)^n=\frac{1}{1+i\omega\tau_R}$, with $|i\omega\tau_R|<1$ indicating the radius of convergence. Causality requires to treat this operator at its entirety, means any finite truncation of the expansion of this operator will violate the mode conserving causality condition \cite{Hoult:2023clg},
\be
{\cal{O}}_{\omega}[F(\omega,\k\neq0)]= {\cal{O}}_{\vert\k\vert}[F(\omega=a\vert\k\vert,\k=\bf{b}\vert{\k}\vert]~.
\label{caus-cond}
\ee
Since these nonlocal operators always bear the risk of divergence, it is indicative to cast them into a local set of equations by raising them to the left hand side of the dispersion equations in a relaxation operator like form. The gradient series \eqref{dispshear}, \eqref{dispdiff},
\eqref{dispsound} show that for each increasing spatial order (higher power of $k$), the power over the relaxation operator also becomes higher, which according to condition \eqref{caus-cond} is essential for causality as well. This is the reason why causal theories need newer degrees of freedom to maintain causality with each higher order of spatial derivative correction. The infinite order correction over temporal derivatives are inevitable in a causal theory and an all order resummation scheme (such as new degrees of freedom in causal M{\"u}ller-Israel-Stewart theory) needs to be incorporated to make the theory consistent with causality. More details can be found in Ref \cite{Bhattacharyya:2024tfj,Mitra:2024yei}.

\section{Conclusion}
In the current analysis, a collision kernel for the relativistic kinetic equation has been proposed in terms of macroscopic field corrections, which conserves both particle current and energy-momentum tensor by construction. The collision term offered here gives a convenient means to study the emerging pole structure  from the kinetic equation. Here, it needs to be mentioned that by the pole structure, simply the singularity of the determinant of $A$ (from matrix equation \eqref{matrix}) has been indicated. Complicated non-analyticity, such as multiple poles or branch cuts, like those originate in the analysis of Green's function singularity \cite{Denicol:2011fa}, are not discussed here.
The poles then offer the dispersion spectra in terms of logarithmic divergences. An expansion of that in the long wavelength limit results in an infinite series in powers of the spatial derivatives  identical to the hydrodynamic gradient series. The crucial observation is that each higher-order spatial gradient term is perfectly balanced by the temporal derivatives appearing in the denominator in relaxation operator-like form. This feature turns out to be essential for the mode conservation and consequently the preservation of the causality for a given theory.

This study reinforces the idea that the microscopic information of the underlying theory manifests itself in some way or other in the emerging hydrodynamic formalism. It could either decide the radius of convergence, setting the scale for the theory \cite{Denicol:2011fa}, or force the theory to include new `non-fluid' degrees of freedom \cite{Bhattacharyya:2024tfj,Mitra:2024yei} other than the fundamental fluid variables. Probably, that is the reason for the riddle that why the causal hydrodynamic derivations never let go off the non-hydrodynamic features, which appear every now and then either in the forms of additional degrees of freedom or in terms of field redefinition of fluid variables.

\section{Acknowledgements}
I duly acknowledge Arpan Das, the discussions with whom laid the foundation of the problem.
For the ﬁnancial support, I acknowledge the Department of Atomic Energy, India.

\end{document}